\documentclass[%
 aip,
 rsi,%
 amsmath,amssymb,
reprint,%
floatfix,%
]{revtex4-1}

\usepackage{graphicx}
\usepackage{dcolumn}
\usepackage{bm}
\usepackage{german, ngerman}
\usepackage[german]{babel}

\usepackage[latin1]{inputenc}
\draft 

\begin{document}
\renewcommand{\figurename}{Fig.} 

\title{Dark Current Measurements on a Superconducting Cavity using a Cryogenic Current Comparator} 



\author{R. Geithner}
\email{rene.geithner@uni-jena.de}
\affiliation{Institut für Festkörperphysik, Friedrich-Schiller-Universität Jena, Helmholtzweg 5, D-07743 Jena, Germany}

\author{R. Neubert}
\affiliation{Institut für Festkörperphysik, Friedrich-Schiller-Universität Jena, Helmholtzweg 5, D-07743 Jena, Germany}

\author{W. Vodel}
\affiliation{Institut für Festkörperphysik, Friedrich-Schiller-Universität Jena, Helmholtzweg 5, D-07743 Jena, Germany}

\author{P. Seidel}
\affiliation{Institut für Festkörperphysik, Friedrich-Schiller-Universität Jena, Helmholtzweg 5, D-07743 Jena, Germany}

\author{K. Knaack}
\affiliation{DESY Hamburg, Notkestrasse 85, D-22607 Hamburg, Germany}

\author{S. Vilcins}
\affiliation{DESY Hamburg, Notkestrasse 85, D-22607 Hamburg, Germany}

\author{K. Wittenburg}
\affiliation{DESY Hamburg, Notkestrasse 85, D-22607 Hamburg, Germany}

\author{O. Kugeler}
\affiliation{Helmholtz-Zentrum-Berlin, Albert-Einstein-Str. 15, D-12489 Berlin, Germany}

\author{ J. Knobloch}
\affiliation{Helmholtz-Zentrum-Berlin, Albert-Einstein-Str. 15, D-12489 Berlin, Germany}

\date{\today}

\begin{abstract}
This paper presents non-destructive dark current measurements of superconducting TESLA cavities for superconducting LINACs. The measurements were carried out in a highly disturbed accelerator environment using a low temperature superconducting d.c. SQUID based Cryogenic Current Comparator (CCC). The overall current sensitivity under these rough conditions was measured to be 0.2 nA/Hz$^{1/2}$ which enables the detection of dark currents of 5 nA.
\end{abstract}

\maketitle 

\section{Introduction}\label{Intro}
Although TESLA (Tera Electron Volt Energy Superconducting Linear Accelerator) cavities were developed for operating in pulsed-mode like in the TESLA linear collider \cite{Brinkmann:2001tech} or the European X-FEL \cite{Altarelli:2006zza} at DESY Hamburg, a lot of planned continuous wave driven LINACs for FEL (free electron laser) will be based on cavities using the TESLA technology like the BESSY FEL. \cite{2004tech}

\begin{figure*}
	\centering
		\includegraphics[width=1.00\textwidth]{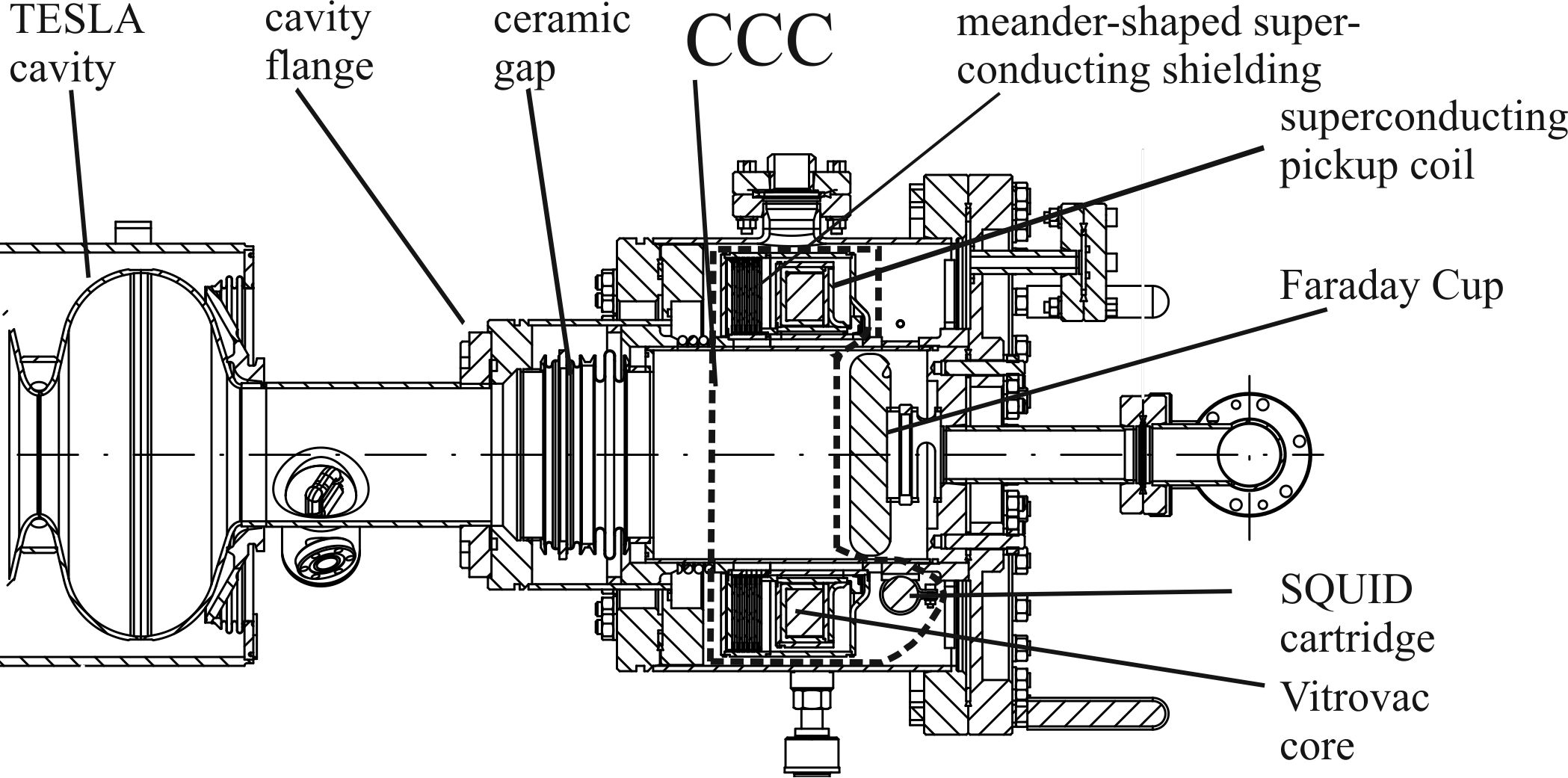}
	\caption{Schematic view of the HoBiCaT test facility with TESLA cavity, CCC and Faraday Cup. The dashed line is added to indicate the CCC with its components.}
	\label{fig:Figure1}
\end{figure*}

Dark currents are one of the limiting factors for cavities with high accelerating gradients. They might cause activation or damage of accelerator components \cite{Collab} and induce quenches in superconducting cavities. Dark currents can excite unwanted higher order modes in the cavity and create a beam halo. They also put an extra load on the cryogenic system. For example, a dark current of 1 $µ$A dumped into in a 20 MV/m cavity will create an extra cryo-load of 20 W. Thus, to analyze the quality of mass-produced cavities the measurement of the absolute value of the dark current depending on the gradient of the acceleration field is necessary. \cite{ISI:000250731700020} Configurations for complete module tests with many cavities will not allow the use of simple Faraday Cups because the energy of the dark current electrons might reach very high energies (some 100 MeV) so that they are not stopped by a Faraday Cup. In this framework a CCC can provide a number of advantages over the measurement of dark currents utilizing a Faraday Cup:
\begin{itemize}
	\item non-destructive measurements,
	\item	measurement of the absolute value of the dark current,
	\item independence of the electron trajectories and energies,
	\item	accurate absolute calibration with an additional wire loop and
	\item	extremely high resolution.
\end{itemize}
The suitability of a CCC as a beam monitor has already been demonstrated, see reference \onlinecite{peters:163}.
Although the presented system was designed for the module test facility of the European X-FEL at DESY Hamburg the measurements were carried out in the Horizontal Bi-Cavity Test-facility (HoBiCaT) at Helmholtz-Zentrum-Berlin (HZB). The HoBiCaT test facility (see Fig. \ref{fig:Figure1}) was constructed to enable rapid-turn-around tests of 9-cell cavities with all the required ancillary. \cite{kugeler:074701}
The CCC is mounted on one end of the cavity with a bellow between. A ceramic gap is recessed in the bellow so that the magnetic field of the dark current can freely propagate through this gap out of the beam pipe to the entrance slit of the CCC (see Fig.~\ref{fig:Figure1}). 
The working temperature of the test facility is 1.8 K by cooling the niobium cavities with superfluid helium.

The presented apparatus senses dark currents in the nA range. The setup also contains a Faraday Cup as a second measurement device for comparison. In the case of the HoBiCaT with one cavity it will work reliably.

\section{Cryogenic Current Comparator}\label{CCC}

Originally developed to compare two currents with one another, \cite{ISI:A1972N887300016} CCCs are used in many different ways in metrology, see for example references \onlinecite{ISI:000165585200035,ISI:000270924900005}.

\begin{figure}
	\centering
		\includegraphics[width=0.45\textwidth]{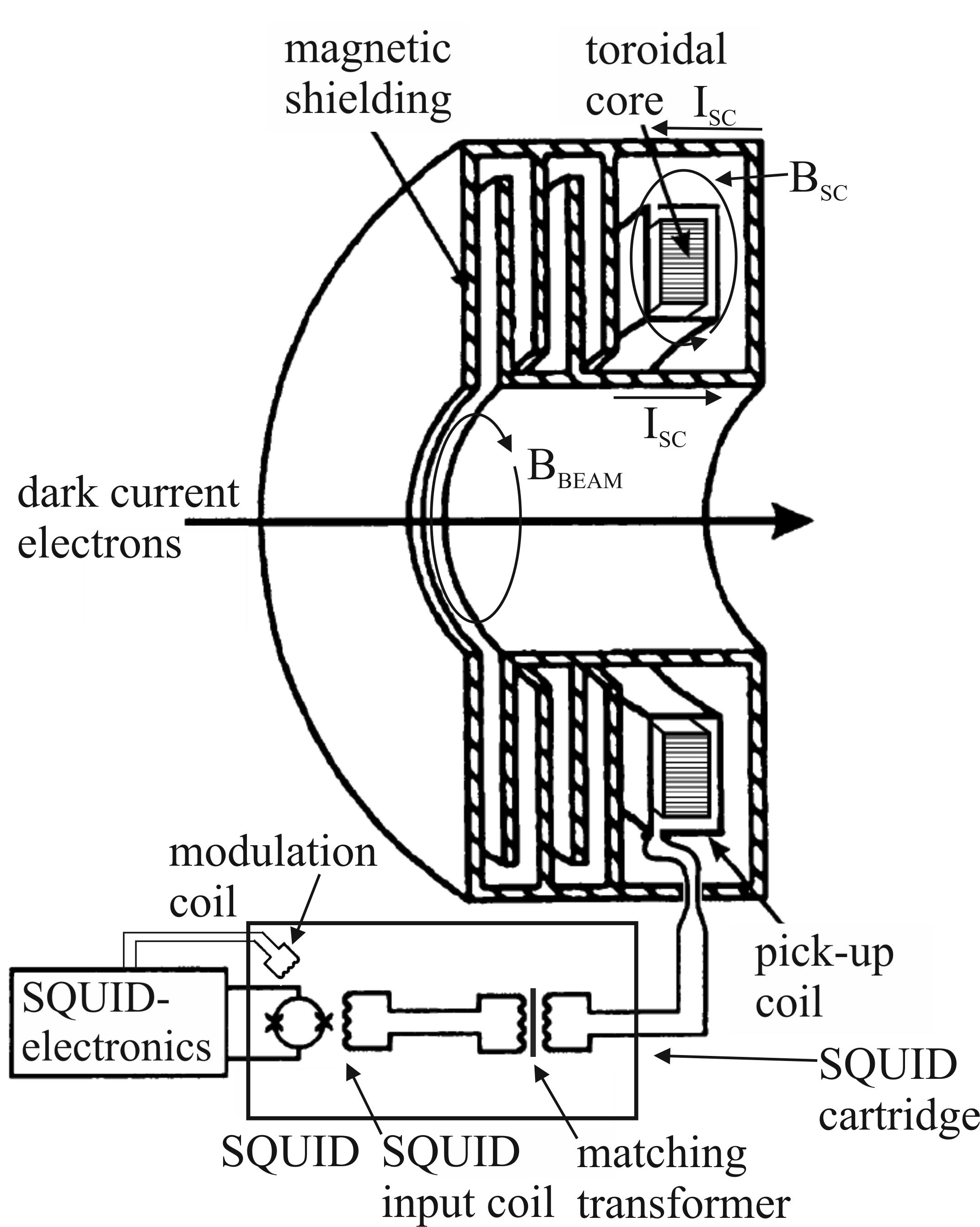}
	\caption{Simplified schematic view of the meander-shaped magnetic shielding, the toroidal pickup coil, the matching transformer and the SQUID.}
	\label{fig:Figure2}
\end{figure}

The presented CCC consists of (see Fig. \ref{fig:Figure2}):
\begin{itemize}
	\item a meander-shaped superconductive niobium shielding,
	\item a superconductive niobium single-turn toroidal pick-up coil with a ferromagnetic core,
	\item high performance d.c. SQUID system.
\end{itemize}

\subsection{Superconductive Shielding}\label{shield}
The resolution of the CCC is reduced if the toroidal pick-up coil operates in presence of disturbing external magnetic fields. In practice, external fields are unavoidable, therefore an extremely effective shielding has to be applied. A circular meander-shaped (\textquotedblleft ring cavities\textquotedblright) superconductive shielding structure (see Fig. \ref{fig:Figure2}) allows passing only the azimuthally magnetic field component of the dark current, while the non-azimuthal field components are strongly attenuated. The attenuation characteristics of CCC shields have been studied analytically in detail in references \onlinecite{Grohmann1976423,Grohmann1976601,Grohmann1977579}. These studies have been applied to the shielding of the presented CCC, and an attenuation factor of approximately 120 dB for the transverse, non-azimuthal magnetic field components is estimated. This result is based on the superposition of the analytic results for the different shielding substructures, in this case coaxial cylinders and \textquotedblleft ring cavities\textquotedblright (as described in detail in reference \onlinecite{Gutmann}).

\subsection{Pick-up Coil}\label{Pick-up Coil}
\begin{figure}
	\centering
		\includegraphics[width=0.45\textwidth]{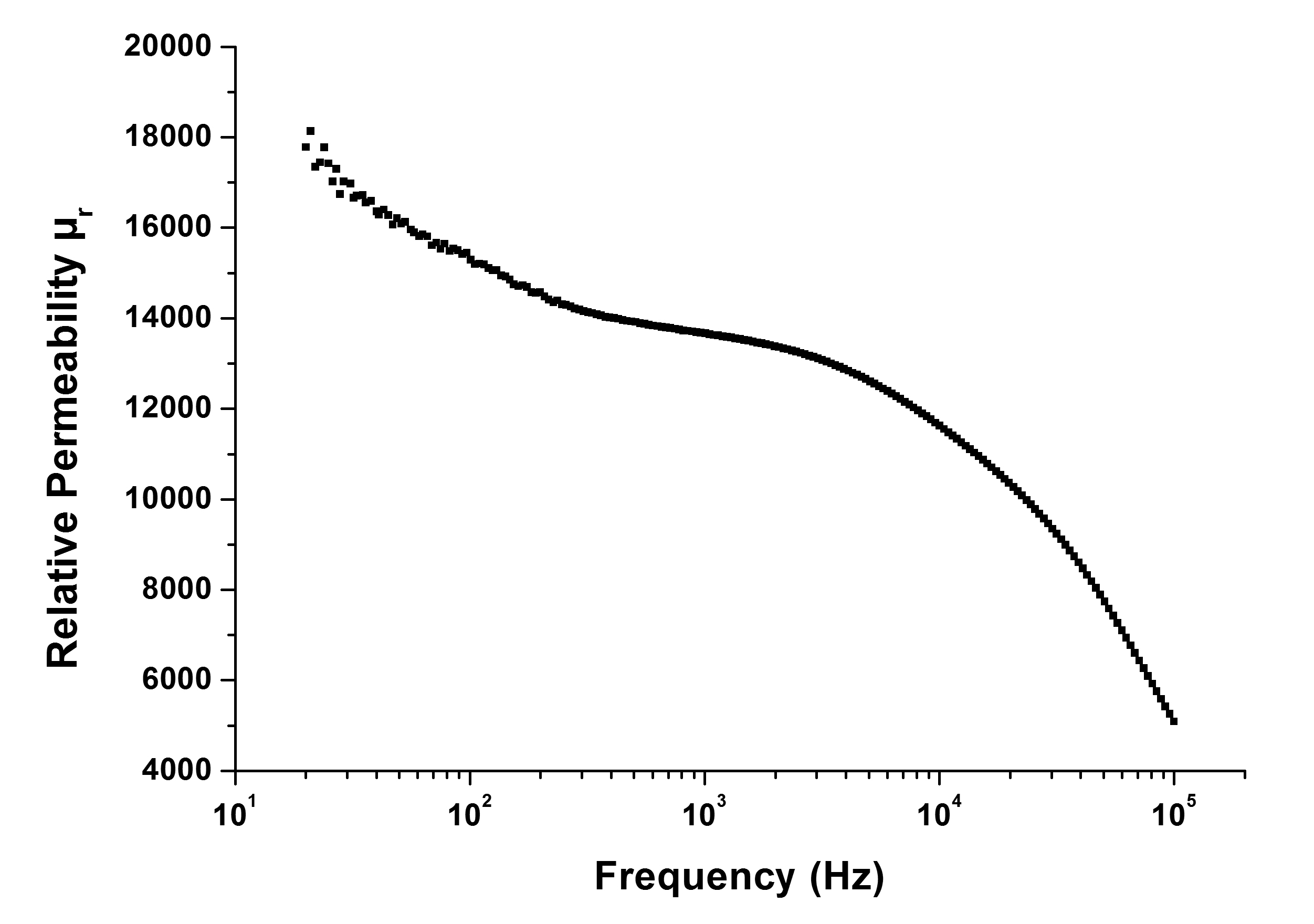}
	\caption{Relative permeability of the ferromagnetic core material Vitrovac 6025F \cite{Vitrovac} at 4.2 K.}
	\label{fig:Figure3}
\end{figure}

A single turn pick-up coil is formed as a superconducting niobium toroid with a slot around the circumference. It contains a Vitrovac 6025F core \cite{Vitrovac} providing a high permeability $µ_{r}$ of about 18000 at liquid helium temperatures (see Fig. \ref{fig:Figure3}). The relative permeability is calculated from the measured inductance of the installed pick-up coil. The measurement was done by an Agilent E4980A Precision LCR Meter; details see reference \onlinecite{Step09}. The material inhomogeneity of the core is averaged by complete encapsulation of a toroidal niobium coil.

\subsection{SQUID System}\label{SQUID}
The key component of the CCC is the high performance d.c. SQUID system developed and manufactured at Jena University. The SQUID sensor UJ 111 \cite{ISI:A1992KB15500007} is designed in a gradiometric configuration and based on Nb-NbOx-Pb/In/Au Josephson tunnel junctions with dimensions of $3 \times 3\:µm^{2}$. The SQUID electronics consists of the low noise preamplifier and the SQUID control unit. The low source impedance of the SQUID (about 1 $\Omega$) is stepped up to the optimal impedance of the preamplifier by a resonant transformer. The d.c. bias and flux modulation (modulation frequency 307 kHz) are fed into the SQUID via voltage-controlled current source situated in the preamplifier and the controller, respectively. An optimal choice of bias and flux modulation working point for the SQUID system was adjusted in the magnetic shielded room (MSR) manufacted by VAC Hanau.\cite{Vitrovac} In a d.c. coupled feedback loop, the field of the dark current to be measured is compensated at the pick-up coil by an external magnetic field generated by the attached electronics. Due to the superconductivity of all leads in the input circuitry (shielding, pick-up coil, transformer, and SQUID input coil) the CCC is able to detect even d.c. currents. For an optimum coupling between the single-turn toroidal pick-up coil (26 $µ$H @ 100 Hz, 4.2 K; relative permeability see Fig. \ref{fig:Figure3}) and the SQUID input coil (0.8 $µ$H) a matching transformer is necessary. Using a modulation frequency of 307 kHz the measurement system provides an over-all bandwidth of 20 kHz (signal level 1 $\Phi_{0}$) or 70 kHz (signal level 0.1 $\Phi_{0}$).

\subsection{Working Principle of the CCC}\label{working}
The dark current electrons creating a magnetic field $B_{BEAM}$ which induces screening currents $I_{SC}$ in the meander-shaped superconductive shielding (see Fig.\ref{fig:Figure2}. These screening currents in turn generate a magnetic field $B_{SC}$ which is transformed into a current through the pick-up coil. This current is then measured by the SQUID, after passing and  being transformed by the matching transformer.

\section{Measurement Setup}\label{Measurement Setup}
The output signals from the CCC and the Faraday Cup were captured by a computerized data acquisition system. The SQUID signal was connected to a National Instruments 
PCI-6221 (M Series DAQ) card. The Faraday Cup was installed at the end of the cavity vacuum chamber (see Fig. \ref{fig:Figure1}). The Faraday Cup was biased with + 210 Volts which was tested to be sufficient to collect all charges (charge vs. bias reached plateau). For its biasing and readout a high resolution DVM was used (Keithley 2400 Source Meter) which was connected via GPIB to the computer to measures the collected dark current to ground after passing the CCC. The signals from the CCC and the Faraday Cup were captured simultaneously in two separate loops with the help of a customized program. The start and the end time as well as a separate measurement time created in each loop, were logged to ensure a chronological correlation. The collected data were analyzed by an applicative tool.
The spectral current noise distributions were measured with a Hewlett Packard 33120A spectrum analyzer.

\section{Results}\label{Results}
The presented measurements of dark currents at the HoBiCaT test facility at HZB were performed on the DESY Z86 cavity. The cavity could be operated either in CW or pulsed-mode. The repetition rate in pulsed-mode was chosen as 0.1 Hz, with 50\% duty cycle.
\begin{figure}
	\centering
		\includegraphics[width=0.47\textwidth]{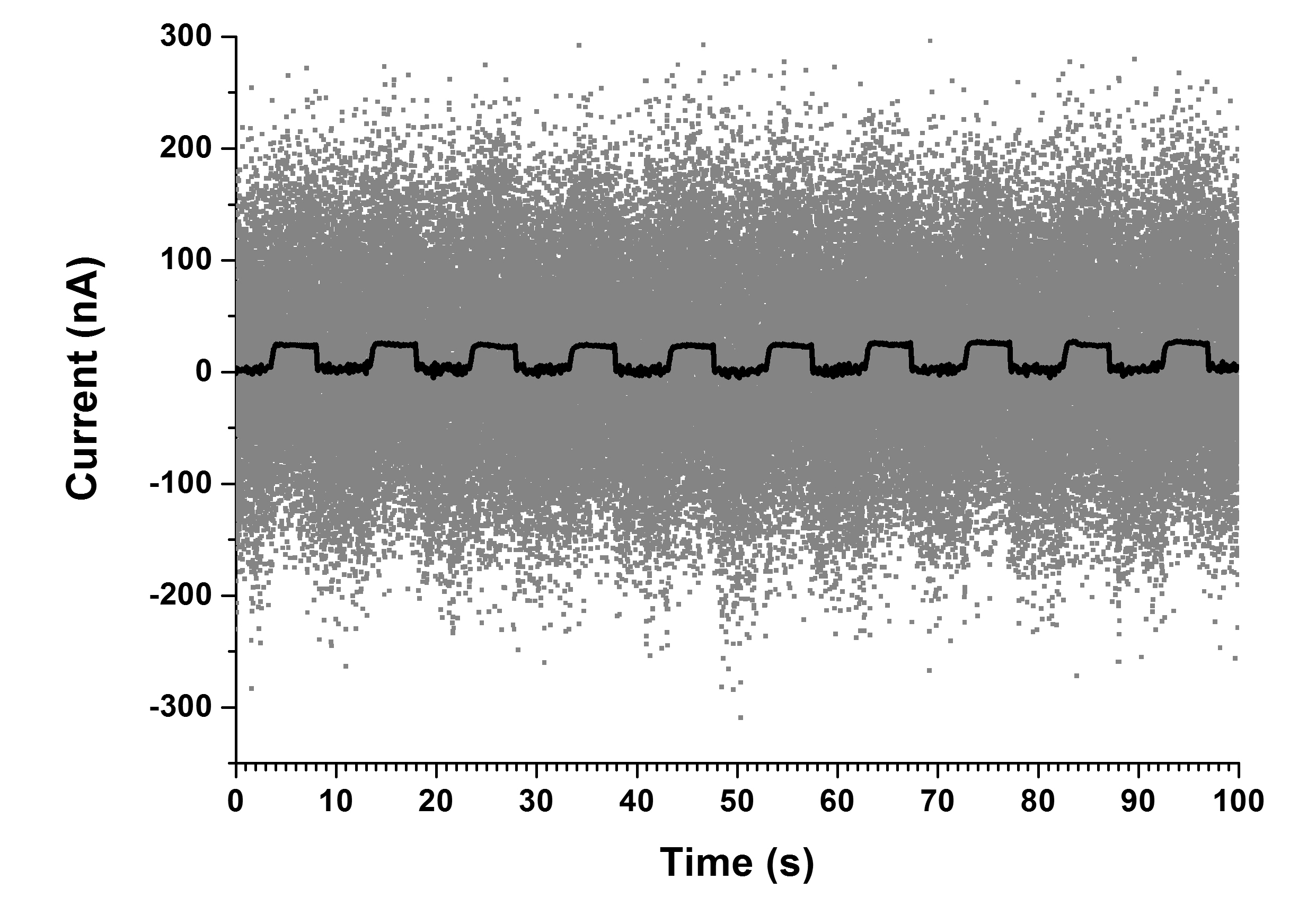}
	\caption{Dark current from the raw CCC-SQUID-signal (grey), calculated with an overall current sensitivity of 243 nA/V within these settings. The black curve is the signal filtered by a 5 Hz low pass.}
	\label{fig:Figure4}
\end{figure}
As depicted in Figure \ref{fig:Figure4} the CCC signal is highly disturbed. Because of the high noise level of more than 1 $\Phi_{0}$ the most sensitive measuring range of the SQUID-electronics could not be used. In this environment an overall current sensitivity of the CCC of 243 nA/$\Phi_{0}$ could be achieved, which means 243~nA/V within these settings. This is slightly smaller compared to Vodel et al., \cite{ISI:000250731700020} where the overall current sensitivity was specified to 200 nA/$\Phi_{0}$. This is due to a necessary reconstruction, which caused a change of the inductance of the pickup coil and therewith a change of the transfer ratio of the matching transformer.
\begin{figure}
	\centering
		\includegraphics[width=0.47\textwidth]{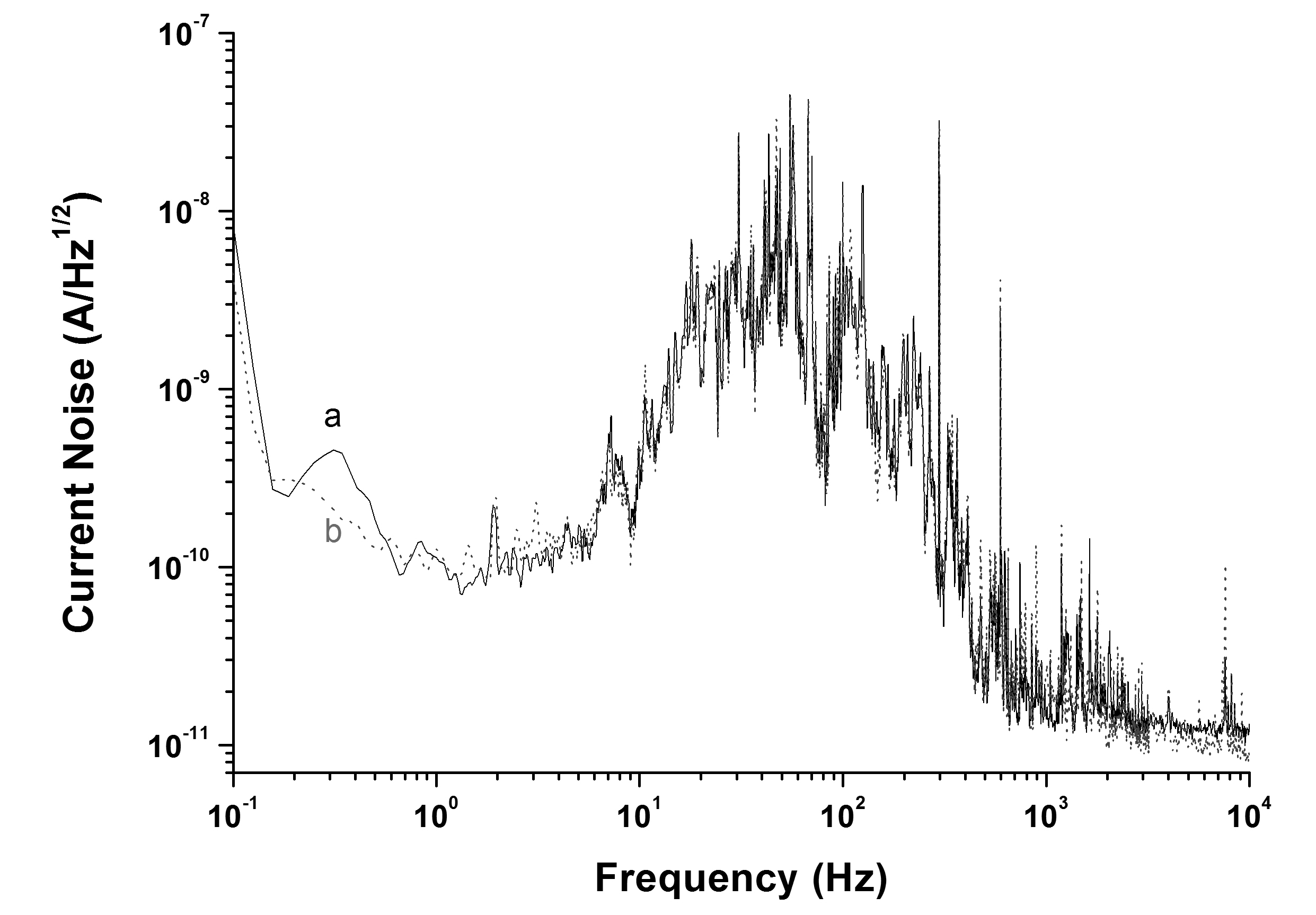}
	\caption{Spectral current noise distributions of the CCC at  HZB (curve a; black) without acceleration field, and with acceleration field (curve b; light grey, dotted).}
	\label{fig:Figure5}
\end{figure}
As one can see in the spectral current noise distributions (see Fig. \ref{fig:Figure5}) there is a considerable noise contribution between 5 Hz and 500 Hz while running the CCC in this disturbed accelerator environment. But it is noticeable that there is no visible additional noise when the acceleration field is switched on. The noise limited current resolution in the frequency range up to 5 Hz is 0.2 nA/Hz$^{1/2}$ with a total noise of 1.8 nA. Between 5 Hz and 500 Hz it increases up to 50 nA/Hz$^{1/2}$ due to external disturbances. A major part of the additional noise seems to be caused by microphonic effects. The peak at 300 Hz in Figure~\ref{fig:Figure5} for example is clearly assignable to the vacuum pump for the isolating vacuum of the HoBiCaT test facility. Neumann et al. \cite{Neumann:10} investigated the microphonic detuning of these cavities in the HoBiCaT test facility regarding the stability of the cavities resonance frequency. They identified several microphonic sources with their corresponding frequencies. One can distinguish between deterministic narrow bandwidth sources, like vacuum, water respectively helium cooling pumps, and random broadband machinery noise. The frequency spectrum of microphonic disturbances was also measured for the DESY Z86 cavity. Compared with the measured flux noise of the CCC (see Fig. \ref{fig:Figure6}) microphonic effects could be confirmed as a source of many disturbances. The coupling of the vibrations of the HoBiCaT test facility to the CCC could be explained by a relative movement of the disks of the meander-shaped shielding against each other.
\begin{figure}
	\centering
		\includegraphics[width=0.45\textwidth]{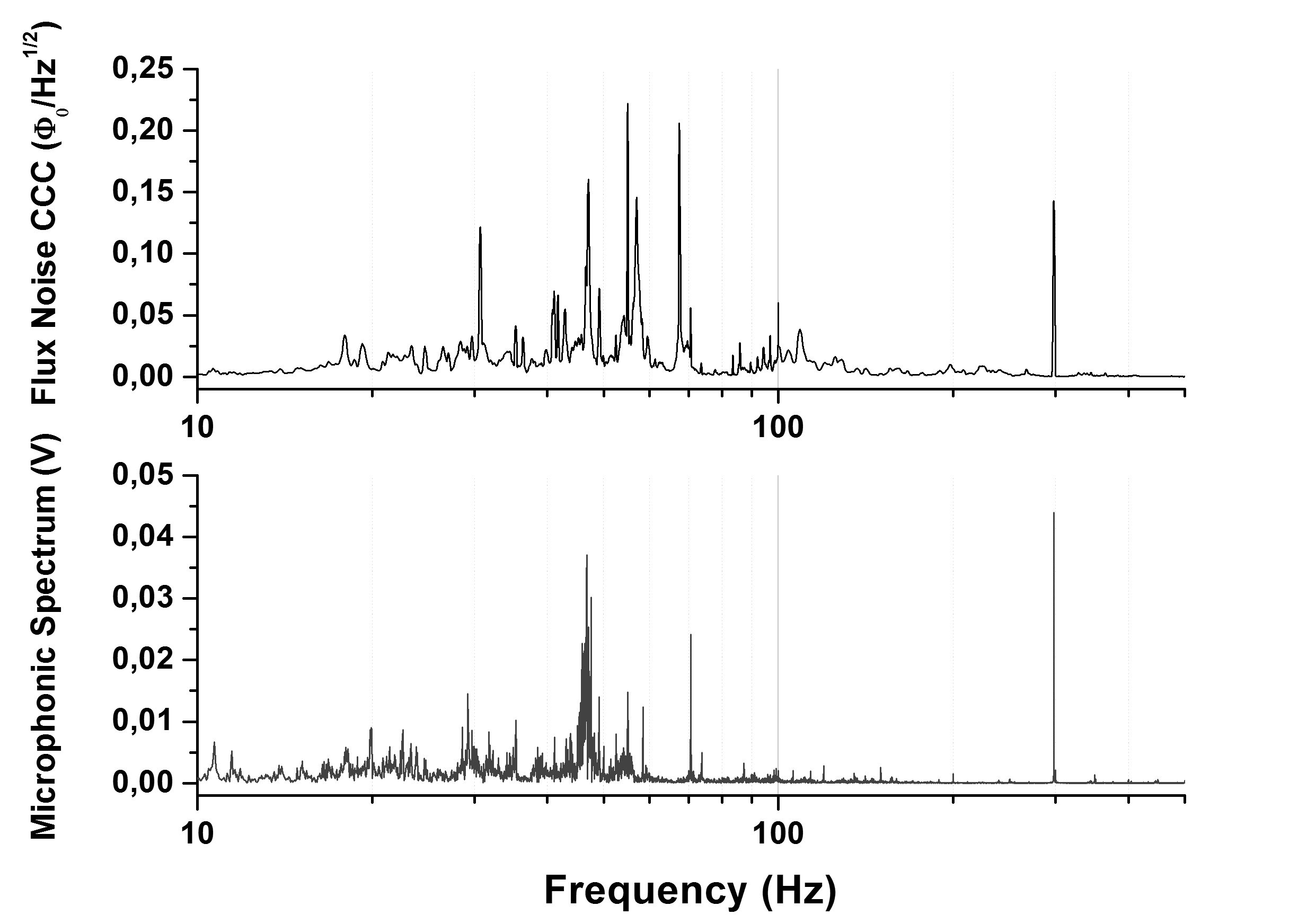}
	\caption{Comparison between spectrum of microphonic disturbances (lower curve) measured by  HZB and flux noise spectrum of the CCC (upper curve).}
	\label{fig:Figure6}
\end{figure}
The pulsed dark current gets clearly visible after filtering the signal with a low pass whose cut-off frequency of 5 Hz is below the additional noise contribution. In Figure \ref{fig:Figure7} a dark current of approximately 21 nA is measured with the Faraday Cup. After filtering and smoothing, the CCC shows similar results. As depicted in Figure \ref{fig:Figure8} it is possible to detect dark currents with an accuracy of several nA down to 5 nA within a 5 Hz bandwidth. Even down to 0.5 nA the pulse sequence is detectable.
The drift in Figures \ref{fig:Figure7} and \ref{fig:Figure8} can be explained by the superconducting properties of the CCC. Since the CCC is sensitive to d.c. and a.c. signals, it also detects slowly varying stray fields. They might be caused by temperature changes, Helium pressure fluctuations or charging of the test facility in pulsed operation mode. A further explanation would be the movement of captured magnetic flux as flux vortices in the superconducting components. It is also striking that the plateau of the dark current when acceleration field is switched on is not constant, but very slightly sloping. This behavior was observed in subsequent measurements at  HZB in a much larger scale. Switching on the acceleration field much longer than 5 sec led to a drop in dark currents of more than 50\%. This was due to a slow heating of higher order modes (HOM) couplers, which were installed in the cavity, so that the quality factor of the cavity had deteriorated slowly. This led in turn to a smaller transmitted power or smaller cavity field and thus to a smaller dark current. The dark current was getting smaller until a balance of HOM-heating and field reduction was achieved. The time constant for this thermal process was several minutes.
\begin{figure}
	\centering
		\includegraphics[width=0.45\textwidth]{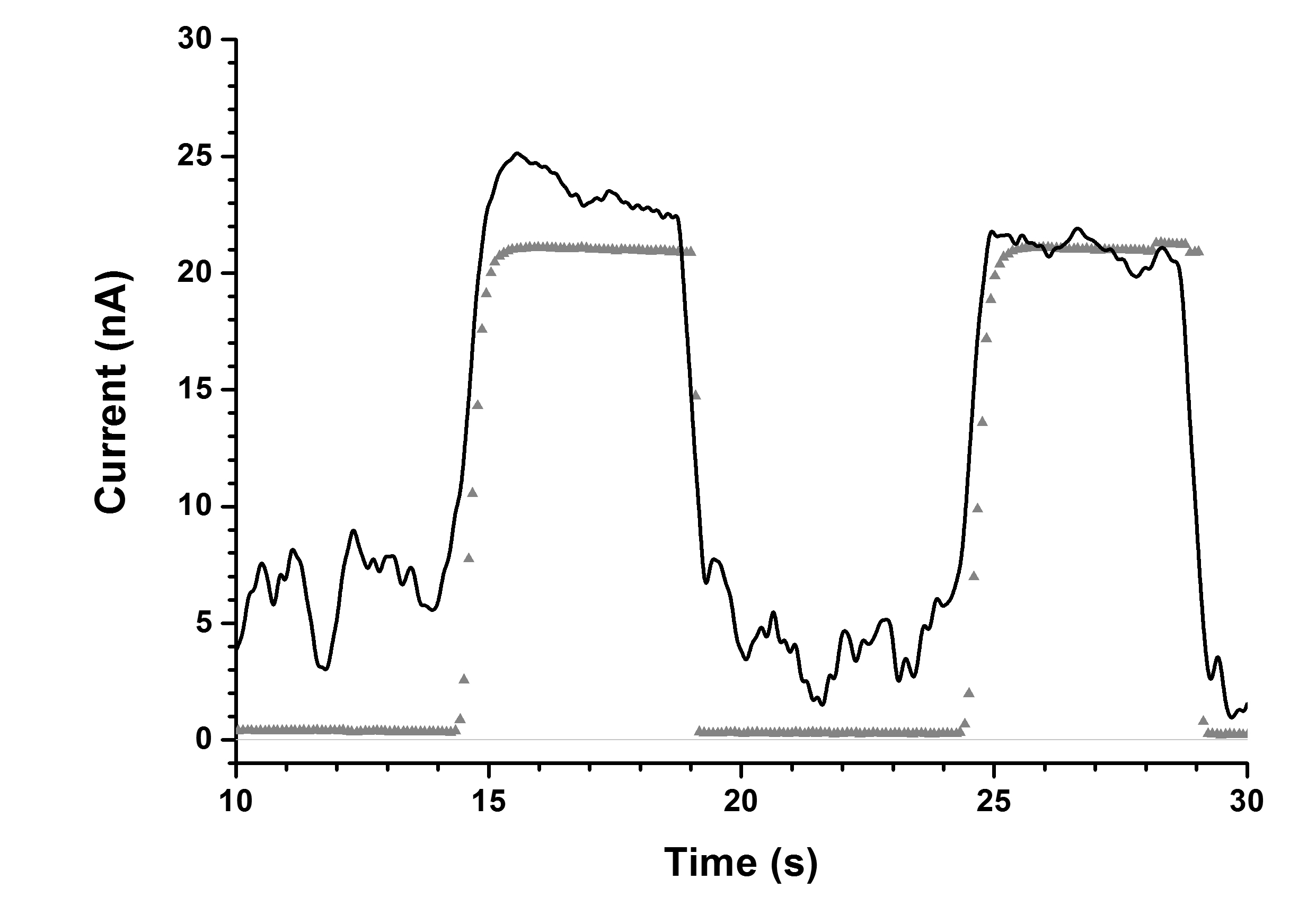}
	\caption{Dark current of approximately 21 nA measured by CCC (black curve, filtered and smoothed) and Faraday Cup (grey, dotted).}
	\label{fig:Figure7}
\end{figure}
\begin{figure}
	\centering
		\includegraphics[width=0.45\textwidth]{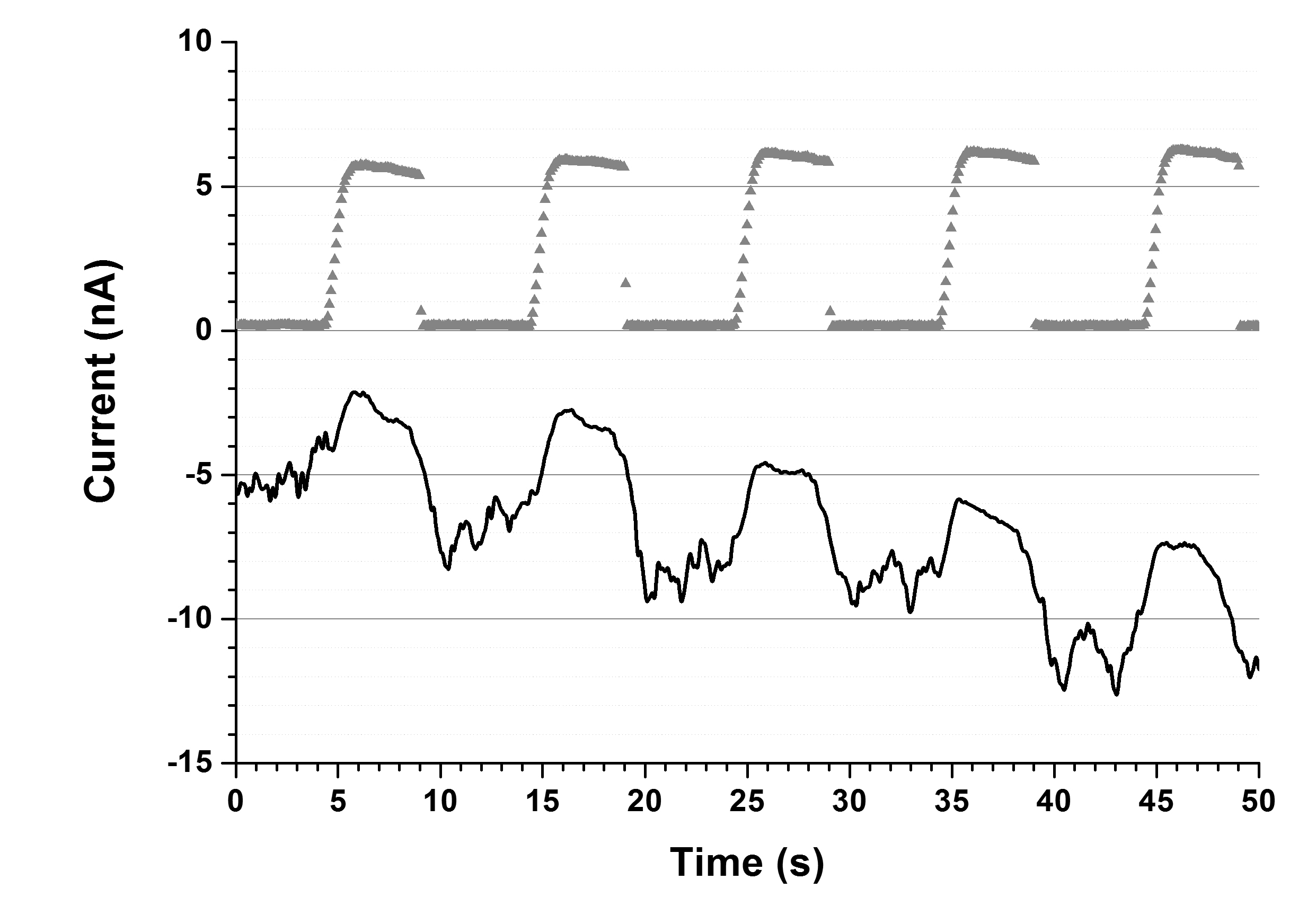}
	\caption{Dark current of approximately 5 nA measured by CCC (black curve, filtered and smoothed) and Faraday Cup (grey, dotted).}
	\label{fig:Figure8}
\end{figure}

\section{Conclusions}\label{Conclusions}
The measurements at the HoBiCaT test facility showed that our Cryogenic Current Comparator is capable of measuring dark currents of superconducting cavities in a highly disturbed accelerator environment.
The overall current sensitivity of the CCC was 243 nA/$\Phi_{0}$ at these measurements. The expected \cite{ISI:000250731700020} noise-limited current resolution of 0.5 nA/Hz$^{1/2}$ was even exceeded. It was measured to be 0.2 nA/Hz$^{1/2}$ for frequencies below 5 Hz. In the frequency range between 5 Hz and 500 Hz it increased up to 50 nA/Hz$^{1/2}$ due to external disturbances, which could be verified in part as microphonic effects. An attenuation of this kind of disturbances might be achieved by fixing the disks of the meander-shaped shielding among each other. Nevertheless the non-destructive detection of dark currents down to 5 nA with an accuracy of a few nA in an accelerator environment was successful.
Further improvements are possible by finding suitable ferromagnetic core materials in terms of absolute value and frequency dependence of the relative permeability. This is part of ongoing investigations; some first results see reference \onlinecite{Step09}.


%
%

%

\begin{acknowledgments}
This work was supported in part by the Deutsches Elektronen
Synchrotron (DESY) Hamburg, Germany. We also thank P. Hanse
as well as the company pro-beam AG, Burg, Germany, and the mechanics crew from the MDI group at DESY for their
expert assistance in manufacturing most of the mechanical
parts for the CCC. Furthermore, we thank HZB for providing the measurement time at the HoBiCaT test facility and their engineers for their help in installation of the CCC.
\end{acknowledgments}

\renewcommand{\refname}{}
\nocite{*}
\bibliography{aipref}

\end{document}